\documentclass[preprint, prfluids, aps, longbibliography
]{revtex4-2}

\usepackage{xcolor} 
\usepackage{graphicx}
\usepackage{dcolumn}
\usepackage{bm, amsmath}

\begin{document}

\title{\textbf{Dimensional regimes in Kolmogorov Flow}}

\author{Melisa Y. Vinograd}
 \affiliation{Universidad de San Andrés, Buenos Aires, Argentina}
  \affiliation{Departamento de Física, Universidad de Buenos Aires, CABA, Argentina}
\author{Joaquín Cullen}%
\affiliation{%
Universidad de San Andrés, Buenos Aires, Argentina}%

\author{Patricio Clark Di Leoni}
 \email{Contact author: pclarkdileoni@udesa.edu.ar}
\affiliation{
Universidad de San Andrés, Buenos Aires, Argentina
}%
\affiliation{
CONICET, Argentina
}%

\date{November 2025}

\newcommand{\CITE}[1]{\textcolor[HTML]{E76F51}{\textbf{[Citation needed]}}}
\newcommand{\MELISA}[1]{\textcolor[HTML]{2A9D8F}{\textbf{[Melisa: #1]}}} 
\newcommand{\PATRICIO}[1]{\textcolor[HTML]{D1495B}{\textbf{[Patricio: #1]}}} 
\newcommand{\JOAQUIN}[1]{\textcolor[HTML]{2780F5}{\textbf{[Joaquín: #1]}}} 

\newcommand{\REVA}[1]{\textcolor{black}{#1}}
\newcommand{\REVB}[1]{\textcolor{black}{#1}}

\newcommand{\Rey}{\mathrm{Re}}
\newcommand{\fref}[1]{\figurename~\ref{#1}}

\begin{abstract}
We study the dimensionality of two-dimensional Kolmogorov flows over a wide range of Reynolds numbers and forcing wavenumbers $k_f=\{2,4,8\}$ using two complementary approaches: convolutional autoencoders and a Kaplan–Yorke estimation based on Lyapunov analysis.
As the Reynolds number increases, two distinct transitions are observed: the first corresponds to the destabilization of a periodic orbit, while the second marks the saturation of the large-scale motions.
When expressed in terms of the forcing Reynolds number, these transitions occur at nearly the same value for all forcing wavenumbers, suggesting a universal scaling with respect to the forcing scale.
By filtering the data to retain only the large-scale range ($k < k_f$), we show that the dimensionality estimated by the autoencoders also saturates at the second transition, implying that once the large scales are fully developed, the subsequent increase in dynamical activity occurs predominantly at smaller scales.
At higher Reynolds numbers, the Kaplan–Yorke dimension ceases to grow, revealing its limited sensitivity to the nonlinear interactions that dominate in this regime.
Both the Kaplan–Yorke saturation dimension and the filtered large-scale dimensionalities exhibit a linear dependence on $k_f$, indicating that the number of active degrees of freedom scales with the forcing scale rather than with the total number of available Fourier modes.
\end{abstract}

\maketitle

\section{Introduction}
The question of what is the number of degrees of freedom in a turbulent flow has a long and rich history. In homogeneous and isotropic flows, Landau~\citep{landau_fluid_2013} proposed the simple, yet powerful, idea that the number of degrees of freedom was equal to the number of vortices needed to fill the whole volume, arriving to the famous $\Rey^{9/4}$ scaling by invoking Kolmogorov theory in three-dimensional flows. Another way to look at the problem comes from dynamical systems theory, where the Kaplan-Yorke conjecture~\citep{kaplan_chaotic_1979} can be used to estimate the size of a chaotic attractor, an approach that has been tried in two-~\citep{grappin_Computation_1987,inubushi_covariant_2012} and three-dimensional ~\citep{tran_Number_2009,hassanaly_lyapunov_2019} turbulence. A third possible way of studying the dimension of the attractor is  through norm inequalities and manifold theory, which can produce upper and lower bounds ~\citep{constantin_Dimension_1988,gibbon_Attractor_1997,robinson_Finitedimensional_1995}. Finally, data-driven approaches, ranging from principal component analysis to autoencoders, also provide a way to assess the number of degrees of freedom of the system~\citep{bruntonBook}. Below we give further information and references on the approaches just listed that are relevant to this work.

In respect to Lyapunov- and dynamical systems-based analysis: \citet{grappin_Computation_1987,grappin_Lyapunov_1991} showed that the Kaplan-Yorke dimension was sensitive to the forcing scale in two-dimensional flows, \citet{yamada_Inertial_1988} showed that in shell models of 2D turbulence all the unstable modes are correlated to those of scales larger than the forcing, and \citet{hassanaly_lyapunov_2019} and \citet{Clark_Tarra_Berera_2020} showed that the estimated dimension may scale slightly slower than what Landau scaling suggests. These studies and others cited above were performed at low Reynolds numbers, due to the computational demands of the calculations involved. Another set of meaningful insights come the study of the Kuramoto-Sivashinky equation, for which the existence of an inertial manifold has been proven~\citep{hyman_kuramoto-sivashinsky_1986}. \citet{yang_hyperbolicity_2009,ding_estimating_2016} showed that a subset of covariant Lyapunov vectors~\citep{ginelli_characterizing_2007} are dynamically decoupled from the rest, owning to the hyperbolicity of the system. They then equate the dimensionality of the system to the number of dynamically active modes, which they deemed ``physical modes''. Interestingly, this estimate was consistently higher than the one obtained via the Kaplan-Yorke conjecture. It is important to note that similar analysis performed on the two-dimensional Navier-Stokes \citep{inubushi_covariant_2012} showed that, except at small Reynolds numbers, the flow did not posses a subset of decoupled covariant Lyapunov vectors, and thus the approach could not be extended to that case.

On the data-driven front, autoencoders~\citep{bruntonBook} have proven remarkably effective at discovering low-dimensional nonlinear embeddings of different datasets~\citep{hinton_reducing_2006, GoodBengCour16}. These architectures learn a nonlinear parametrization of the attractor directly from data, effectively acting as a data-driven realization of an inertial manifold. \citet{linot_deep_2020,linot_data-driven_2022} showed that estimates obtained via autoencoders matched those obtained by Lyapunov analysis mentioned above. Extending these ideas to other systems, \citet{Vinograd_Clark} studied the Rayleigh number dependence of the dimension of 2D Rayleigh-Bénard flows, where they showed that the dimension increases abruptly when the flow starts to transition to turbulence. Furthermore, they developed a criteria to determine proper estimates from autoencoders in large systems, revealing that regularization effective for the Kuramoto–Sivashinsky equation do not generalize straightforwardly. In the context of Kolmogorov flow, autoencoders have been coupled with Fourier decomposition
\citet{page_revealing_2021, Page_Holey_Brenner_Kerswell_2024} and proper orthogonal decomposition \cite{kelshaw_Proper_2024} to identify and classify important dynamical structures in the flow. More recently, \cite{cleary_estimating_2025} estimated the dimensionality of 2D Kolmogorov flows as a function of the Reynolds number using autoencoders with implicit rank minimization. They found that the attractor dimension scales as $D \sim \Rey^{1/3}$, a much weaker dependence than the theoretical upper bounds that scale as $\Rey^{4/3}$ \citep{constantin_Dimension_1988} and that the autoencoder embeddings display a rich latent structure, including distinct classes of high-dissipation events and dynamically irrelevant unstable periodic orbits.

Most existing studies have focused either on Lyapunov-based or on data-driven approaches to dimensionality estimation, without establishing a direct link between the two. Here, we aim to bridge this gap by directly comparing autoencoder-based estimates with those obtained from the Kaplan–Yorke conjecture, thus connecting data-driven and dynamical-systems perspectives on flow dimensionality. As a testbed, we consider two-dimensional Kolmogorov flow, a paradigmatic system for studying the transition to turbulence and the statistical properties of two-dimensional flows~\citep{Boffetta_Ecke_2012}. In this geometry, energy injected at the forcing scale undergoes an inverse cascade toward larger scales, while enstrophy is transferred to smaller ones~\citep{kraichnan_inertial_1967}, leading to a rich coexistence of coherent vortices and small-scale fluctuations. \REVA{In practice, the observation of a fully developed dual cascade with well-defined inertial ranges requires a large separation of scales, typically achieved through forcing at sufficiently high wavenumbers and, in some cases, the inclusion of large-scale friction. The simulations considered here do not aim at realizing an asymptotic inverse-cascade regime.} 

Because of its simplicity, well-controlled forcing, and periodic geometry, Kolmogorov flow provides an ideal setting for probing how the effective dimensionality of turbulence depends on the Reynolds number and the forcing scale. Here we study how dimensionality changes with forcing wavenumber and over a range of Reynolds numbers wider than previously reported, and explain the different regimes observed.

The work is organized as follows. Section~\ref{sec:setup} presents the problem set-up and describes the Kolmogorov flow configuration in~\ref{subsec:kolmogorov}, the autoencoder-based framework for dimensionality estimation in~\ref{subsec:AE}, the Lyapunov analysis used for comparison in~\ref{subsec:lyapunov}, and the details of dataset generation and neural network training in~\ref{subsec:data}. Section~\ref{sec:results} reports the main results, including the evolution of the estimated dimensionality with Reynolds number, the identification of distinct dynamical regimes, and the effect of varying the forcing wavenumber on the system. Finally, Section~\ref{sec:conclusions} summarizes the main findings and discusses their implications.

\section{Problem set-up and methodology}
\label{sec:setup}

\subsection{Kolmogorov flow}
\label{subsec:kolmogorov}
In two dimensions, the Navier-Stokes equation for an incompressible flow reads

\begin{equation}
\partial_t \omega + \mathcal{J}(\psi,\omega) = \nu \nabla^2 \omega + (\nabla \times \bm{f})_z,
\label{eq:kolmogorov}
\end{equation}
where $\omega = \partial_x v - \partial_y u$
is the out-of-plane vorticity field
$\omega = \partial_x v - \partial_y u$, associated with the velocity $\bm{u} = (u,v)$.  
 $\psi$ is the streamfunction, and 
$\mathcal{J}(\psi,\omega) = \partial_x \psi\,\partial_y \omega - \partial_y \psi\,\partial_x \omega$ 
is the Jacobian operator representing nonlinear advection, and $\nu$ is the kinematic viscosity.
the Kolmogorov flow is defined by an external monochromatic forcing in the $x$--direction
\begin{equation}
\bm{f}(x,y) = f_0 \sin(k_f y)\,\bm{e}_x,
\end{equation}
where $f_0$ is the forcing amplitude and $k_f$ its wavenumber, with the typical choice being $k_f=4$. Taking $L$ to be the characteristic size of the domain, the Reynolds number is usually defined as

\begin{equation}
Re = \frac{\bigl(L/2\pi\bigr)^{3/2}\,\sqrt{f_0}}{\nu}.
\end{equation}
In order to take into account the forcing of the system we also define a forcing Reynolds number
\begin{equation}
Re_f = \frac{\bigl(L_f/2\pi\bigr)^{3/2}\,\sqrt{f_0}}{\nu}
= Re \, \left(\frac{L_f}{L}\right)^{3/2},
\label{eq:Re_f}
\end{equation}
where $L_f = 2\pi/k_f$ is the forcing lengthscale.  

Two-dimensional incompressible flows exhibit an inverse cascade of energy, where energy flows from $k_f$ to smaller wavenumbers, and a direct cascade of enstrophy, where enstrophy flows from $k_f$ to higher wavenumbers \citep{Boffetta_Ecke_2012}.
The viscous lengthscale, where viscous effects dominate and the enstrophy cascade terminates, can be estimated from the balance between enstrophy flux and viscous dissipation, its inverse takes the form 

\begin{equation}
\ell_\nu^{-1} = k_\nu \sim 
\nu^{-1/2}\Omega^{1/6},
\label{eq:k_nu}
\end{equation}
where $\Omega = \nu \langle |\nabla\omega|^2 \rangle$ is the entrosphy dissipation rate. 

The Kolmogorov flow exhibits several continuous and discrete symmetries.
The governing equation~\eqref{eq:kolmogorov} is invariant under continuous translations in the $x$--direction due to the periodicity and homogeneity of the forcing along this axis.
In the $y$--direction, the sinusoidal forcing introduces a discrete translation symmetry of period $L_f = 2\pi/k_f$.
Additionally, the system is invariant under a reflection about the midplane of the forcing, combined with a change in the sign of the streamfunction.
These symmetries generate multiple dynamically equivalent states connected by translations or reflections.
To avoid redundant representations of the same physical state, we perform a symmetry reduction based on these invariances prior to training the neural networks, as described in Section~\ref{subsec:data}.

\subsection{Autoencoders for dimensionality estimation}
\label{subsec:AE}

The goal of determining the dimension of the manifold in which the solutions of the Kolmogorov flow live can be expressed as that of mapping a large set of realizations of the flow to a lower dimensional space. For that goal we can use an autoencoder.
An autoencoder consists of two components: an encoder $\mathcal{E}_\theta$ and a decoder $\mathcal{D}_\phi$. 
The encoder maps an input field of size ${N \times N}$, in this case the vorticity field $\omega$, to a latent representation 
\begin{equation}
z = \mathcal{E}_\theta(\omega) \in \bm{R}^d,
\end{equation}
where $d \ll N^2$ defines the size of the bottleneck. 
The decoder then attempts to reconstruct the original field from this compressed representation,
\begin{equation}
\hat{\omega} = \mathcal{D}_\phi(z).
\end{equation}
The model parameters $(\theta, \phi)$ are optimized to minimize the mean-squared reconstruction error
\begin{equation}
\mathcal{L}(\theta,\phi) = \frac{1}{N^2} \sum_{i,j=1}^{N}
\left( \omega_{ij} - \hat{\omega}_{ij} \right)^2.
\end{equation}
where this error function is averaged over many realizations of the flow.

In order to estimate the minimum number of latent dimensions needed to properly reconstruct the fields we train separate autoencoders with different values of $d$ and assess their performance. Following previous work \citep{Vinograd_Clark}, we evaluate the scale-by-scale error by comparing the vorticity spectra of both the original and the reconstructed fields. As $d$ is increased more and more scales are properly resolved (in the sense that the error at a given scale is similar or lower than the energy at that scale).
The smallest latent dimension~$d^\ast$ is then identified as the minimum~$d$ for which the resolved wavenumber reaches the viscous (dissipation) wavenumber~$k_\nu$, indicating that all dynamically relevant scales of the flow are accurately reconstructed. Illustrative examples of this criterion, including the corresponding spatial reconstructions and spectral comparisons, are provided in Appendix~\ref{app:reconstruction_criteria}. Details on the implementation, and network architectures and training protocol used are given below.
It is worth mentioning too that while regularized variants where the sweep in $d$ can be omitted have been proposed and found to be useful in systems such as Kuramoto-Sivashinky, they fail in larger, more multiscale systems \citep{Vinograd_Clark}. Therefore, we only work with the non-regularized version just presented. Below we compare our results with previously reported estimates obtained with implicit rank minimization autoencoders \cite{cleary_estimating_2025}.

\subsection{Lyapunov analysis}
\label{subsec:lyapunov}

Lyapunov exponents quantify the exponential growth or decay of infinitesimal perturbations along a trajectory, providing a measure of predictability and dynamical complexity.  
Positive exponents correspond to unstable directions, while negative ones indicate contracting directions.  
Their cumulative sum defines the Kaplan–Yorke dimension~\citep{kaplan_chaotic_1979,kuznetsov_attractor_2020},
\begin{equation}
d^\dag = j + \frac{\sum_{i=1}^{j} \lambda_i}{|\lambda_{j+1}|},
\end{equation}
where $j$ is the largest index satisfying $\sum_{i=1}^{j}\lambda_i \ge 0$.  
The quantity~$d^\dag$ estimates the number of effectively active degrees of freedom on the attractor and serves as a benchmark against which to compare the minimum latent-space dimension~$d^\ast$ obtained from the autoencoders.  

To compute the Lyapunov spectrum of the Kolmogorov flow, we employ a finite-time Lyapunov exponent (FTLE) framework based on the Benettin algorithm~\citep{benettin_lyapunov_1980,edson_lyapunov_2019}.  
This method estimates the finite-time Lyapunov exponents by repeatedly evolving a set of orthonormal perturbations along the trajectory and reorthonormalizing them after each integration interval~$T$.  
The approach considers a state vector~$\bm{U}(t)$ that encodes the velocity field in a one-dimensional arrangement. The flow-map~$\Phi^T$ denotes the nonlinear evolution operator that advances the state vector over a time interval~$T$, and its corresponding Jacobian,~$J=D\Phi^T(\bm{U})$, describes the linearized evolution of infinitesimal perturbations.  
Its action on a perturbation~$\delta\bm{U}$ is approximated numerically by finite differences as
\begin{equation}
    J\,\delta\bm{U} \approx 
\frac{\Phi^T(\bm{U}+\epsilon\,\delta\bm{U}) - \Phi^T(\bm{U})}{\epsilon},
\end{equation}
with~$\epsilon= \sqrt{\epsilon_{\mathrm{mach}}}\,\Vert\bm{U}\Vert / \Vert\delta\bm{U}\Vert$, where $\epsilon_{\mathrm{mach}}$ is the machine precision and $\Vert\cdot\Vert$ denotes the $L_2$ norm.  
This perturbation factor offers a compromise between round-off and truncation errors~\citep{viswanath_recurrent_2007}.  
At each reorthonormalization step, the propagated perturbations are factorized using QR decomposition and the diagonal entries of~$R$ accumulate the instantaneous expansion rates.  
The long-time average of the logarithms of these factors, $\ln(R_{ii})/T$, yields the Lyapunov spectrum~\citep{Ott_2002}, ordered as~$\lambda_1 \ge \lambda_2 \ge \cdots \ge \lambda_N$, where~$N$ is the number of exponents computed.  
The corresponding Kaplan–Yorke dimension~$d^\dag$ provides an estimate of the attractor dimension at each Reynolds number.  
Details on the numerical implementation and the choice of~$T$ are provided in Section~\ref{subsec:data}.


\subsection{Autoencoder, dataset and simulation details}
\label{subsec:data}

All autoencoders share the same architecture: an encoder with four convolutional layers having $[32,\,64,\,128,\,256]$ filters, kernel size $5\times5$, and stride~2, mirrored by a symmetric decoder with transposed convolutions. 
Each layer uses rectified linear unit (ReLU) activations except for the final output layer, which employs a sigmoid activation. 
Networks are trained using the Adam optimizer with a batch size of~16 and early stopping. 
For some configurations, the learning rate was adaptively reduced using a \texttt{ReduceLROnPlateau} scheduler to improve convergence. 
For each Reynolds number, an ensemble of replicas was trained for each latent dimension~$d$ to account for variability, with controlled perturbations in the initialization, train/test splits, and learning-rate schedule. 
The smallest dimension~$d^{\ast}$ that consistently satisfies the predefined reconstruction criteria across the ensemble is reported as the effective dimension. The uncertainty in the reported values arises from the discretization of the tested~$d$ values and the observed variability among ensemble replicas.

To train the autoencoders, we ran numerical simulations across a wide range of Reynolds numbers and for different forcing wavenumbers.
The simulations were performed in a periodic domain $[0, 2\pi)\times[0, 2\pi)$ discretized with $256\times256$ grid points. 
For the highest Reynolds numbers, the resolution was increased to $512\times512$ for the simulations, although the fields were later downsampled to $256\times256$ for the autoencoder analysis (this did not affect the physical content of the fields as the simulations require to be dealiased by the 2/3 rule already). 
The domain size $L=2\pi$ and the forcing amplitude $f_0=1$ were kept fixed, only $\nu$ was varied in order to change the Reynolds number (for each $k_f$). 
The chosen values of $\mathrm{Re}$ span regimes ranging from simple periodic solutions to fully developed turbulent and chaotic behavior. We carried out simulations at $k_f=2$, $4$ and $8$ using the pseudospectral code GHOST~\citep{Mininni_ghost}.
For each case, we performed long integrations exceeding $10^{4}$ turnover times, 
starting from different random initial conditions, 
and saved between one and three vorticity fields per turnover time. The initial transient period was discarded and not included in the datasets used for training and testing.
This procedure yielded datasets on the order of $10^4$ snapshots, 
of which $20\%$ were withheld for testing. 
The turnover time was defined as 
$\tau = L / U_{\mathrm{rms}}$, 
so that for $k_f=4$ the values spanned $\tau_f \in [0.3,\,1.5]$ time units 
across the considered $\mathrm{Re}$. 
Each dataset was normalized to the range~$[0,\,1]$ according to the maximum 
vorticity magnitude in each case, to better accommodate the neural-network 
architecture. Following previous works~\citep{linot_deep_2020,cleary_estimating_2025}, 
we reduced the symmetries present in Kolmogorov flow (see Section~\ref{subsec:kolmogorov}) to obtain more compact data representations and avoid training redundant weights associated with translated or reflected states. The continuous translational symmetry was treated using the method of slices~\citep{budanur_tutorial_2015, Budanur_Cvitanović_Davidchack_Siminos_2015}, 
while discrete shift-reflect and rotational symmetries were handled following the approach described by~\cite{De_Jesus_Linot_Graham_2023} for $k_f=2$, extended here to $k_f=4$ and $8$, following~\cite{cleary_estimating_2025}. 

The Lyapunov exponents were computed using the open-source Python library \texttt{spookyflows}~\citep{spookyflows_2025}, which provides tools for the numerical analysis of chaotic dynamical systems, including Lyapunov spectrum estimation and dimensionality diagnostics.  
Since finite-time Lyapunov exponents depend on the integration time~$T$, we selected~$T$ sufficiently long to capture the divergence of trajectories but not so long as to lose correlation with the initial state.  
A range of integration times was tested, and the final reported values correspond to the average over the interval where convergence was observed.  


\section{Results}
\label{sec:results}

\begin{figure}
    \centering
    \includegraphics[width=0.7\linewidth]{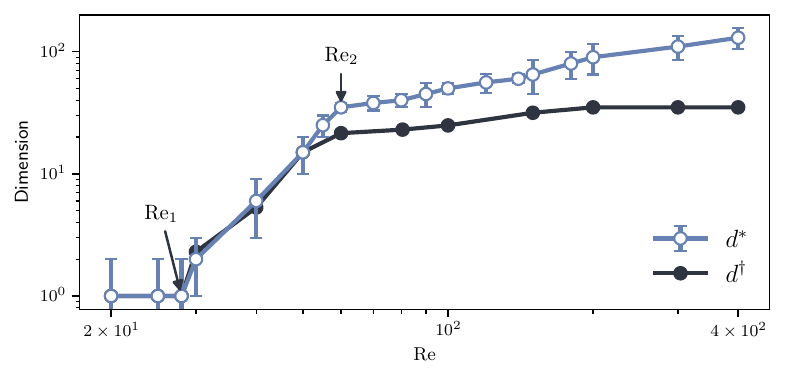}
    \caption{Minimum latent dimension $d^*$ and Kaplan-Yorke dimension $d^\dag$ as function of the Reynolds number $\mathrm{Re}$ for $k_f=4$.}
    \label{fig:d_vs_Re_kf4}
\end{figure}

\subsection{Dimensionality across Reynolds numbers}

We begin by showing results for $k_f=4$, the typical choice in Kolmogorov flows. First we show the dimensionality estimates over a range of Reynolds numbers and then we give further details on the dynamics and our explanations of the observed regimes below. \figurename~\ref{fig:d_vs_Re_kf4} shows the minimum latent dimension $d^*$, obtained by training different autoencoders at the various Reynolds numbers as explained above, and the Kaplan-Yorke dimension $d^\dag$ as function of the Reynolds number. Three distinct regimes can be observed. At low Reynolds number, both estimates are basically equal to 1. It is important to remember that we have removed all symmetries of the flow through phase-shift procedures as mentioned in Sec.~\ref{subsec:data}. Also, the Lyapunov spectra for $\mathrm{Re}=20$, 25 and 28 contain only negative exponents, accordingly, we set $d^\dag \approx 1$ as the minimum physically meaningful dimension. After $\Rey = 28$, which we denote as $\Rey_1$, the estimates begin to increase at a similar pace \REVA{and have comparable values}. At approximately $\Rey=60$, which we denote as $\Rey_2$, \REVA{another transition occurs after which $d^*$ continues to grow but at a slower pace, while $d^\dag$ appears to slowly reach a plateau.} Note too that the change in behavior in $d^\dag$ happens slightly sooner than for $d^*$.  Both estimates are an order of magnitude smaller than the degrees of freedom predicted by the classical Landau argument based on the number of Fourier modes up to the viscous cutoff $k_\nu$, implying that the effective dynamics evolve on a much lower-dimensional manifold. As we explain below, the Kaplan-Yorke dimension is capturing the baseline non-linear behavior of the flow, while the minimum latent dimension can also capture small-scale dynamics. 

\begin{figure}
    \centering
    \includegraphics[width=0.8\linewidth]{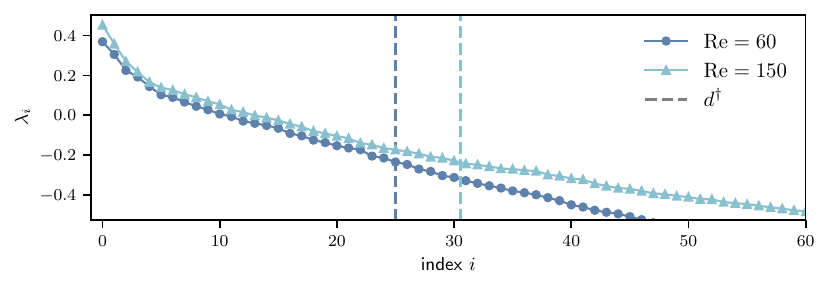}
    \caption{Lyapunov spectra $\lambda_i$ vs.\ index $i$ for two Reynolds numbers, ordered from largest to smallest. Only the leading exponents are shown. Dashed vertical lines mark their respective Kaplan–Yorke dimensions $d^\dag$.}
    \label{fig:CLV_lambdai}
\end{figure}

It is interesting to view these results in light of previously reported ones by \cite{yang_hyperbolicity_2009,ding_estimating_2016} for the Kuramoto-Sivashinsky equation. In that case, it was found that the Kaplan-Yorke dimension was lower than the estimates of the dimension of the inertial manifold obtained via a physical mode analysis (which \cite{linot_deep_2020} later showed to be equal to estimates obtained via autoencoders). In order to look for the possible existence of physical modes we show the Lyapunov spectra for $\Rey=60$ and $\Rey=150$ in \figurename~\ref{fig:CLV_lambdai}. Unlike the Kuramoto-Sivashinky case there is no clear sign of an entangled cluster of eigenvalues separated from all other strongly contracting, hyperbolically isolated directions.
Instead, the exponents decrease smoothly and cross zero without a gap or drop-off. The plateau in $d^\dag(\mathrm{Re})$ therefore reflects a balance between weakly unstable and contracting directions, whereas $d^\ast$ continues to grow beyond the second transition, capturing the increasing state-space complexity even when $d^\dag$ has saturated.
This is consistent with results reported by \cite{inubushi_covariant_2012}, where they showed a lack of hyperbolicity in Kolmogorov flows at high Reynolds numbers.


Before moving on to the analysis of the different regimes, we want to comment on recent results reported in \cite{cleary_estimating_2025}. There the authors present estimates of the dimensionality of Kolmogorov flows obtained via IRMAE \cite{linot_data-driven_2022}, a regularized autoencoder approach which omits the sweep in the bottleneck dimension $d$ required by autoencoders with fixed latent dimension. The resulting estimates show a dependence on Reynolds number that is qualitatively very similar to the one reported here, but with considerably higher values for the dimension. For example, at $\Rey \approx 60$ they report values of the dimension above 200, while ours are around 30. This discrepancy is consistent with previous results focusing on Rayleigh-Bénard flow \cite{Vinograd_Clark}, where it was shown that IRMAE consistently overestimated the autoencoders with fixed bottleneck dimension, or even failed to properly capture the multiscale dynamics. It is both impressive and promising though how both techniques capture the same change in regime at $\Rey=\Rey_2$, giving further backing to autoencoder-based dimensionality estimation techniques. \REVA{Note also that the data shown in \cite{cleary_estimating_2025} hints at the transition at $\Rey=\Rey_1$ but is not directly discussed. The remaining goals of this work are to understand the origin of these transitions and to establish links between $d^\dagger$ and $d^\ast$.}


\begin{figure}[t]
    \centering
    \includegraphics[width=0.9\linewidth]{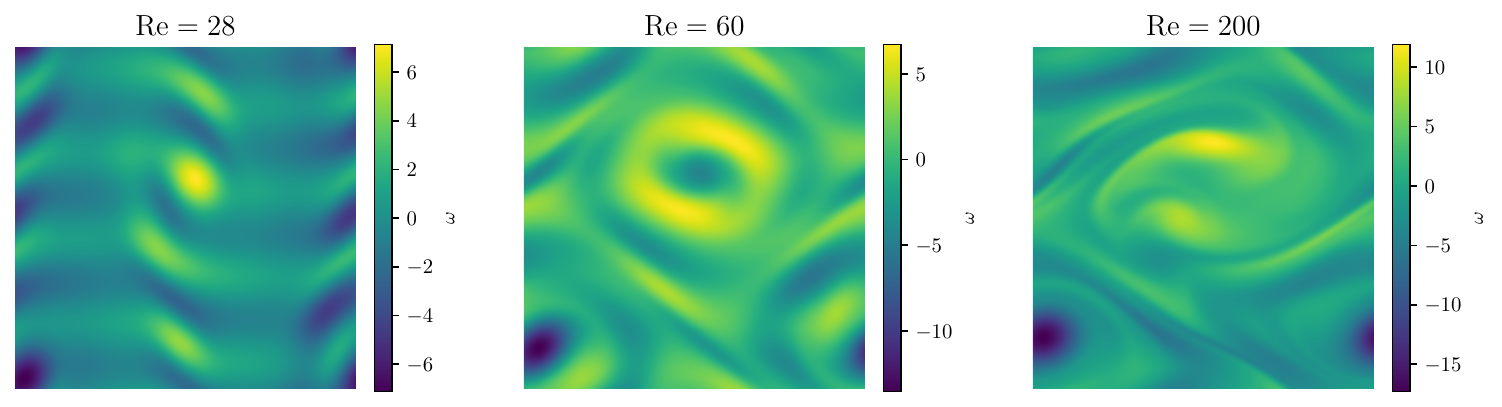}
    \caption{Vorticity fields for Kolmogorov flows at $k_f=4$ and $\mathrm{Re}=28$, $60$, and $200$. Each snapshot uses its own color scale (colorbars are not shared).}
    \label{fig:snapshots_dissipation_kf4}
\end{figure}

\subsection{Analysis of the different dimensionality regimes}

\begin{figure}
    \centering
    \includegraphics[width=0.8\linewidth]{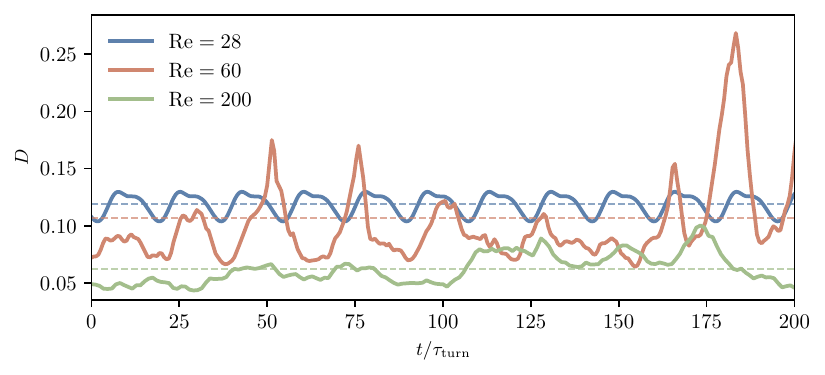}
    \caption{
    Dissipation $D$ for Kolmogorov flow at $k_f = 4$ and $\mathrm{Re} = 28,\,60,\,200$. Time evolution of $D$ in units of turnover times; horizontal dashed lines mark the time-averaged values $\langle D \rangle_t$ for each case.}
    \label{fig:D_timeseries}
\end{figure}

To illustrate the qualitative changes across regimes, \figurename~\ref{fig:snapshots_dissipation_kf4} shows representative vorticity snapshots for Kolmogorov flow at $k_f=4$ at three Reynolds numbers: one at the first transition $\Rey_1$, one at the second transition $\Rey_2$, and one well within the turbulent regime at $\Rey=200$. In the first regime the flow is quasi-periodic and dominated by a pair of large-scale rolls. After the transition at $\Rey_1$ the rolls distort and interact more strongly, and by $\mathrm{Re}=200$ the fields exhibit sharper gradients and elongated filaments, indicative of finer-scale vorticity structures. Complementing these snapshots, \figurename~\ref{fig:D_timeseries} presents time series of the dissipation $D = \nu \langle \omega^2 \rangle$ made dimensionless by the turnover time $t/\tau_{\mathrm{turn}}$ for the same set of Reynolds numbers. Horizontal dashed lines indicate the time-averaged values $\langle D\rangle$ for each case. For the low-$\mathrm{Re}$ case, the $D(t/\tau_{\mathrm{turn}})$ signal at $\mathrm{Re}=28$ is nearly periodic with modest amplitude around $\langle D\rangle$, consistent with the roll-dominated state observed near the first transition. $\mathrm{Re}\!\approx\!60$ displays the most pronounced intermittent excursions, with sporadic bursts rising well above the mean, as reported by \cite{cleary_estimating_2025}, whereas the higher Reynolds number shown here ($\mathrm{Re}=200$) fluctuates more narrowly around its average.

\begin{figure}
    \centering
    \includegraphics[width=0.6\linewidth]{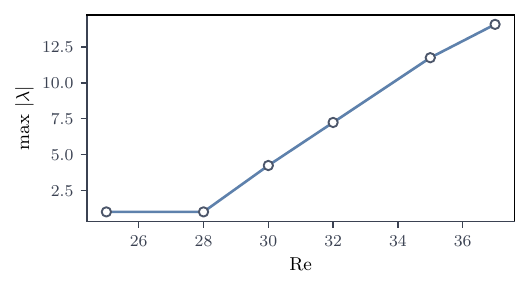}
    \caption{Maximum Floquet multiplier $\max|\lambda|$ versus $\mathrm{Re}$ for the same orbit family.}
    \label{fig:UPO_floquet}
\end{figure}

We now turn our attention to the mechanisms driving the transitions between regimes. As in the first regime the flow is close to a periodic orbit,
the first transition at $\Rey_1$ can be interpreted through the lens of unstable periodic orbits (UPOs). We first located a candidate UPO at that $\mathrm{Re}$ via a recurrence search that monitors the mismatch between snapshots separated by a trial period $T$. This candidate was then converged to a true periodic solution using a Newton--GMRES--Hookstep scheme \cite{viswanath_critical_2008,viswanath_recurrent_2007,Chandler_Kerswell_2013}, which solves the extended Newton system with GMRES and constrains the update within a trust region via a hookstep projection.
Once a UPO is available, its stability is quantified with Floquet multipliers. We continued this same UPO in the parameter $\mathrm{Re}$—using each converged solution as the initial guess at the next $\mathrm{Re}$—and computed the spectrum along the branch. \figurename~\ref{fig:UPO_floquet} shows the maximum modulus $\max|\lambda|$ versus $\mathrm{Re}$. For $\mathrm{Re}\le 28$, all nontrivial multipliers satisfy $|\lambda|\le 1$ (the trivial time-shift multiplier is $\lambda=1$), indicating neutral stability. Beyond this point, $\max|\lambda|$ increases approximately linearly with $\mathrm{Re}$, signaling loss of stability of the periodic solution and aligning with the first transition $\mathrm{Re}_1\simeq 25$–$30$, where the minimum latent dimension $d^\ast$ departs from $1$.

\begin{figure}
    \centering
    \includegraphics[width=0.7\linewidth]{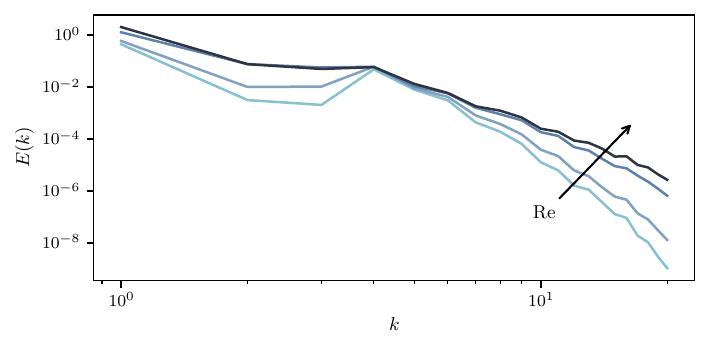}
    \caption{Energy spectra $E(k)$ for $k_f=4$ at Reynolds numbers $\Rey=30,\,40,\,60,$ and $100$. 
The arrow indicates the direction of increasing $\Rey$. 
For $\Rey \geq 60$ the spectra saturate before the forcing scale $k_f=4$, reflecting the buildup of energy at large scales.}
    \label{fig:kf4_spectra}
\end{figure}

During the second regime, the flow becomes increasingly nonlinear, activating more and more scales. This is evidenced in the vorticity spectra shown in
\figurename~\ref{fig:kf4_spectra}: as $\Rey$ increases, the spectra shift upwards both at large and small scales. The spectra also give an indication of what is happening in the second transition at $\Rey_2$. For $\Rey>\Rey_2$ the large scales saturate and the energy contained in those modes stops increasing with $\Rey$.
It is worth pointing out that under certain conditions two-dimensional flows can generate condensates \cite{Boffetta_Ecke_2012}. While this does not seem to be the case for this flow at these Reynolds numbers, it is still evident how the flow is forming large coherent structures.

Returning our focus to dimensionality, the change in behavior observed at $\Rey_2$ in \fref{fig:d_vs_Re_kf4} can be interpreted as the point by which all large scales have been activated.
After this transition, the large-scale pattern is established and vortices become well developed; further increasing $\Rey$ mainly populates the flow with new small-scale features that appear as folding and stretching of existing filaments. One way to visualize this fine-scale roughness in the velocity field is the magnitude of the $\|\nabla^{4}\bm{u}\|$, which weights high wavenumbers (via $k^4$) and is closely connected to palinstrophy (e.g. $\langle \bm{u}\!\cdot\!\nabla^{4}\bm{u}\rangle = \|\nabla^{2}\bm{u}\|_2^2 = 2P$ in 2D, under periodic/no-flux conditions). As $\Rey$ grows, $\|\nabla^{4}\bm{u}\|$ reveals the intensification and thinning of elongated filaments, indicating increasingly pleated small-scale structure.

\begin{figure}
    \centering
    \includegraphics[width=0.7\linewidth]{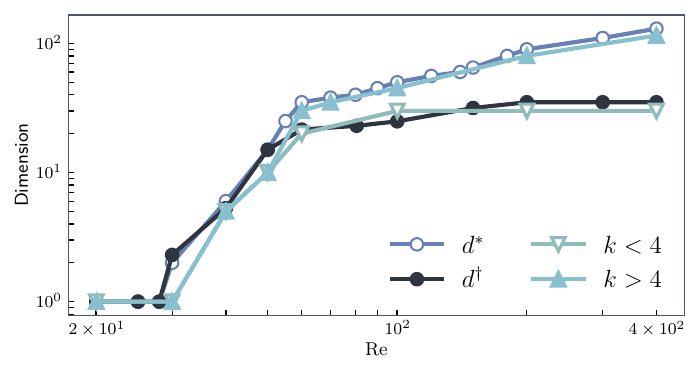}
    \caption{Minimum latent dimension $d^\ast$ versus Reynolds number $\Rey$ split by wavenumber bands. Contribution from modes $k<4$ and contribution from modes $k>4$. Grey open circles show the total (“Full”) from the full field for reference and the band indicates the range of the Kaplan–Yorke dimension $d^\dag$, both already shown in \figurename~\ref{fig:d_vs_Re_kf4}. The growth of $d^\ast$ at higher $\Rey$ is dominated by the $k>4$ modes.}

    \label{fig:Nmas_Nmenos}
\end{figure}

Beyond $\Rey_2$, the Kaplan–Yorke dimension $d^\dag$ saturates, while the autoencoder–inferred dimension $d^\ast$ continues to grow. We propose that this separation reflects two facets of complexity: $d^\dag$ counts the number of effectively unstable (linear) directions via the Lyapunov spectrum, whereas $d^\ast$ is sensitive to additional, predominantly nonlinear degrees of freedom \REVA{that do not correspond to new positive Lyapunov exponents}. In short, $\Rey_2$ marks the completion of linear–mode activation; \REVA{beyond this point, no additional unstable directions are created in phase space}, while the increase in complexity captured by $d^\ast$ is mostly nonlinear, and $d^\dag$ remains approximately constant. This is in line with what \citet{yamada_Inertial_1988} observed for shell models, where all unstable Lyapunov vectors corresponded to modes larger than the forcing. \REVA{More generally, these results support the view that the Kaplan–Yorke dimension is dominated by large-scale dynamics, with inertial-range and small-scale modes contributing mainly non-positive Lyapunov exponents.} It is also consistent with recent synchronization experiments \cite{Inubushi_Caulfield_2025}, that showed that information only up to the forcing scale is needed in order to synchronize Kolmogorov flows, \REVA{highlighting that large-scale degrees of freedom are sufficient to control the chaotic dynamics}. To solidify and give further evidence to this argument, we analyze the contributions of the large and small scales to the total dimensionality. To do this we first split each dataset into large- and small-scale snapshots by applying a sharp spectral filter at the forcing wavenumber $k_f=4$, and then train separate autoencoders on (i) fields retaining only modes $k<k_f$ and (ii) fields retaining only modes $k>k_f$. \REVA{This procedure allows us to isolate, in a controlled manner, the degrees of freedom associated with the large-scale dynamics captured by the Kaplan–Yorke dimension from those associated with smaller-scale variability.} For the $k < k_f$ networks, a bottleneck size~$d$ was considered sufficient when $E_\Delta(k) \ll E(k)$ for all retained modes $k < k_f$. For the $k>k_f$ networks, we used the same criterion as for the full fields, requiring the error spectrum to cross the reference spectrum at $k>k_\nu$. The resulting minimum dimensions $d^\ast(\Rey)$ for each filter are shown in \figurename~\ref{fig:Nmas_Nmenos}. Note that the sum of the two filtered dimensions is slightly larger than the full-field value, which is expected since there is shared information. The large-scale contributions ($k<4$) follow the Kaplan–Yorke dimension closely and also reach a plateau after $\Rey_2$, whereas the small-scale contributions ($k>4$) continue to grow. \REVA{This wavenumber-resolved analysis therefore provides a mechanistic link between the dynamical-systems perspective embodied by the Kaplan–Yorke dimension and the data-driven autoencoder-based estimate, showing explicitly that $d^\dag$ captures the large-scale component of the flow complexity.}

\begin{figure}[t]
    \centering
    \includegraphics[width=0.9\linewidth]{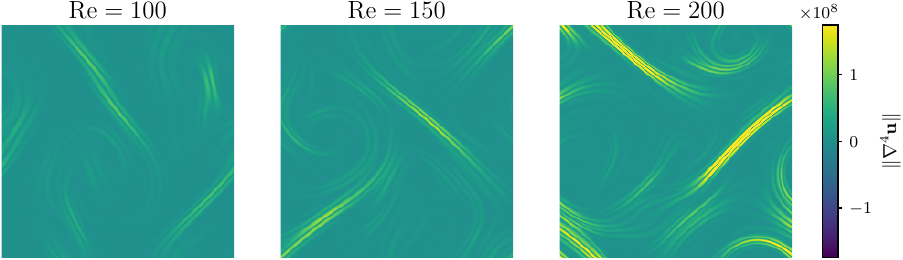}
    \caption{Instantaneous magnitude of $\|\nabla^{4}\bm{u}\|$, for Kolmogorov flow forced at $k_f=4$ at three Reynolds numbers ($\Rey=100,\,150,\,200$). The panels share a common color scale.}
    \label{fig:nabla4u_triptych}
\end{figure}

\subsection{Effects of varying forcing wavenumber}

\begin{figure}
    \centering
    \includegraphics[width=0.9\linewidth]{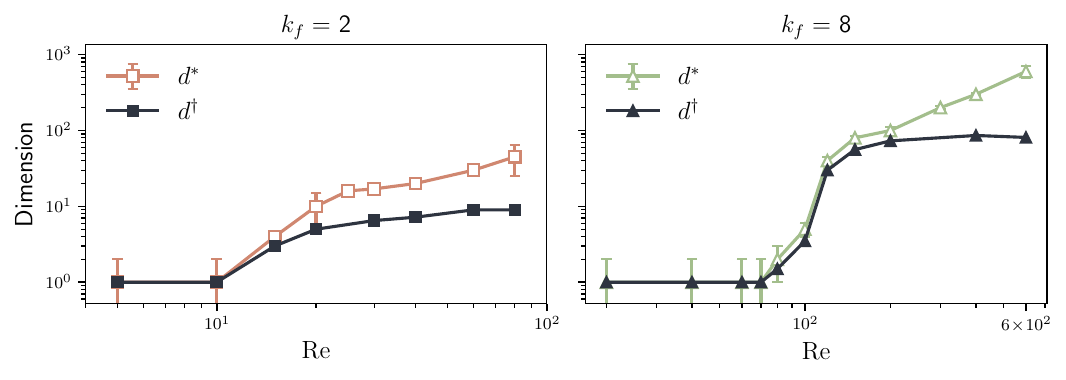}
    \caption{
    Minimum latent dimension $d^{\ast}$ and Kaplan--Yorke dimension $d^{\dag}$ as functions of the Reynolds number for different forcing wavenumbers~$k_f$.
    \textbf{Left:} $k_f=2$. \textbf{Right:} $k_f=8$.}
    \label{fig:dstar_ddag_kf2_kf8}
\end{figure}

\begin{figure}
    \centering
    \includegraphics[width=0.9\linewidth]{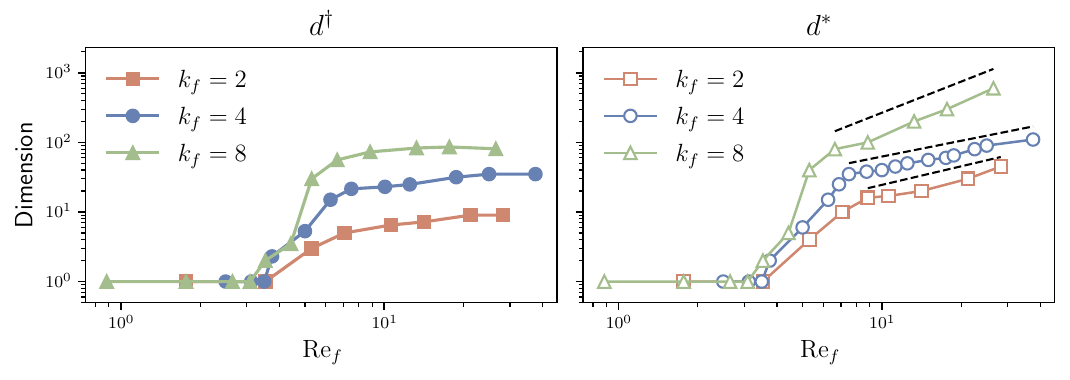}
    \caption{Dependence of the estimated dimension on the forcing Reynolds number~$\Rey_f$ for different forcing wavenumbers~$k_f = 2, 4, 8$. 
    \textbf{Left:} Kaplan-Yorke dimension~$d^{\dagger}$ computed from the Lyapunov spectrum. 
    \textbf{Right:} minimum latent dimension~$d^{\ast}$ obtained from the autoencoder reconstructions. \REVA{Dashed lines correspond to power-law fits of the form 
    $d^\ast \sim \Rey_f^{\alpha}$, with fitted exponents 
    $\alpha \approx 0.9$, $0.75$, and $1.5$ for $k_f = 2$, $4$, and $8$, respectively.}}
    \label{fig:ddag_vs_dstar_Ref}
\end{figure}

We now explore the effects of varying the forcing scale $k_f$.
\figurename~\ref{fig:dstar_ddag_kf2_kf8} shows the comparison between the $d^\ast$ and $d^\dag$ for $k_f = 2$ and $k_f = 8$.
Both cases exhibit the same patterns reported for $k_f = 4$ in \figurename~\ref{fig:d_vs_Re_kf4}:
the system is first locked on to a single orbit, after a first transition the dimensionality starts to increase, with $d^\ast$ and $d^\dag$ having very similar values. After a second transition $d^\ast$ continues to grow at a slower rate while $d^\dag$ plateaus. A better comparison between the three cases is presented in \figurename~\ref{fig:ddag_vs_dstar_Ref}, where the dimensionality estimates obtained for each $k_f$ are shown as a function of the forcing Reynolds number $Re_f$, as defined in~\eqref{eq:Re_f}. Both the first and second transitions are now aligned between the three cases, giving further credence to the dimensional interpretation of these transitions.

\REVA{In Fig.~\ref{fig:ddag_vs_dstar_Ref} we show empirical power-law fits of $d^\ast$ as a function of $\Rey_f$ for each forcing wavenumber $k_f$, yielding exponents of approximately $0.9$, $0.75$, and $1.5$ for $k_f=2$, $4$, and $8$, respectively. 
We emphasize, however, that the available scaling ranges are limited and the uncertainties in the dimensional estimates remain significant; consequently, we cannot make any strong claims about the scaling behavior and the values are presented mostly for reference.}

\begin{figure}
    \centering
    \includegraphics[width=0.7\linewidth]{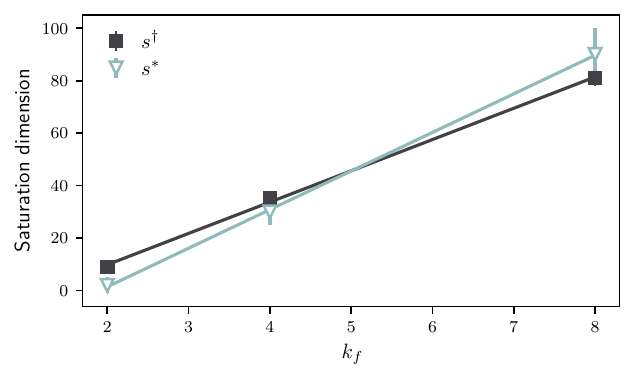}
    \caption{
        Saturation dimension as a function of the forcing wavenumber~$k_f$.
        Squares indicate the plateau value of the Kaplan--Yorke dimension,~$s^{\dagger}$,
        obtained as the saturation value reached with increasing Reynolds number.
        Triangles correspond to~$s^{\ast}$, the minimum latent dimension of the autoencoders
        trained on datasets containing only modes with~$k \leq k_f$,
        where the value of~$d^{\ast}$ saturates with increasing~$\Rey$.
        Solid lines show linear fits to each quantity, revealing that both
        scale approximately linearly with~$k_f$. The slopes are approximately
        11.9 for~$s^{\dagger}$ and 14.7 for~$s^{\ast}$.
    }
    \label{fig:DKY_Nless_vs_kf}
\end{figure}

Finally, we are interested in understanding how the dimensionality varies a function of $k_f$. For that we define the saturation value $s^\ast$ as the $d^*$ obtained by analyzing low-pass-filtered fields at high $\Rey$ (as in \figurename~\ref{fig:Nmas_Nmenos}) and
$s^\dag$ as the mean $d^\dag$ over the plateaus. 
The values of $s^\ast$ and $s^\dag$ for each $k_f$ are shown in \figurename~\ref{fig:DKY_Nless_vs_kf}. Quite surprisingly, both saturation values scale linearly in $k_f$, instead of quadratically, as could be expected in a two-dimensional system. Linear fits are also displayed: the slope is 11.9 for the saturation of the Kaplan–Yorke dimension~$s^{\dagger}$, and 14.7 for the saturation obtained from autoencoders trained with data filtered to $k < k_f$, $s^{\ast}$. Although the slope for the autoencoder-based estimate is larger, the saturation values remain within the uncertainty of the data points. At first glance, we cannot offer a clear explanation to this behavior, and, as stated below, we will explore it further in future work.

\section{Conclusions}
\label{sec:conclusions}

In this work, we calculated two estimates of the dimensionality of two-dimensional Kolmogorov flows over a wide range of Reynolds numbers and for different forcing wavenumbers. One estimate was obtained from convolutional autoencoders, while the other was based on the Kaplan–Yorke dimension. These two estimates were found to be in close agreement, indicating their robustness and validity.
The variation of the dimensionality with Reynolds number revealed two distinct transitions in the dynamics. The first transition corresponds to the loss of stability of a periodic orbit, while the second arises from the saturation of the large-scale motions in the flow.
Moreover, we observed that the two transitions occur at approximately the same value of the forcing Reynolds number~$Re_f$ for all forcing wavenumbers considered, suggesting a universal scaling of the flow organization with respect to the forcing scale.

The Kaplan–Yorke dimension agrees closely with the autoencoder-based estimate before the second transition but saturates afterward, indicating that it does not capture the full range of nonlinear interactions and smaller-scale structures activated at higher~$Re$.
By filtering the data to isolate the large-scale range ($k < k_f$), we found that the dimensionality estimated by the autoencoders also saturates at the second transition, indicating that beyond this point the large-scale structures are fully developed and the subsequent increase in dynamical activity occurs predominantly at smaller scales.
We defined the corresponding saturation dimension as that of the large scales. Interestingly, this saturation dimension scales linearly with the forcing wavenumber, indicating that a one-to-one correspondence between Fourier modes and degrees of freedom is not possible, since the number of Fourier modes grows quadratically with~$k_f$.
Future work will explore different forcing configurations to further investigate this scaling behavior and to clarify the activation of nonlinear interactions across the turbulent transition.

\begin{acknowledgments}
This work was partially supported by the Google PhD Fellowship Program.
\end{acknowledgments}
 Code is available at \url{https://github.com/melivinograd/Kolmogorov-Autoencoder}. 

\appendix
\section{Appendix}

\subsection{Reconstruction criteria}
\label{app:reconstruction_criteria}

In this appendix, we provide additional details on the reconstruction criteria introduced in Section~\ref{subsec:AE}, where the autoencoders were employed to estimate the effective dimensionality of the flow. 
For each network with bottleneck size~$d$, we compute the original vorticity spectrum~$Z(k)$ and the error spectrum~$Z_\Delta(k)$ obtained from the reconstruction error $\Delta = \omega_z - \tilde{\omega}_z$, where $\tilde{\omega}_z$ is the autoencoder output. The comparison between $Z(k)$ and $Z_\Delta(k)$ allows us to assess which wavenumbers are faithfully reconstructed: the flow is considered resolved up to the largest wavenumber for which the error energy remains significantly smaller than the flow energy, beyond which the two spectra become comparable. The circular markers in the spectra indicate these limiting wavenumbers~$k_r(d)$, representing the smallest scales still accurately reconstructed. We define $d^\ast$ as the smallest latent dimension whose autoencoder resolves all scales up to the dissipation wavenumber~$k_\nu$ for the corresponding~$\Rey$.

\begin{figure}
    \centering
    \includegraphics[width=1.0\linewidth]{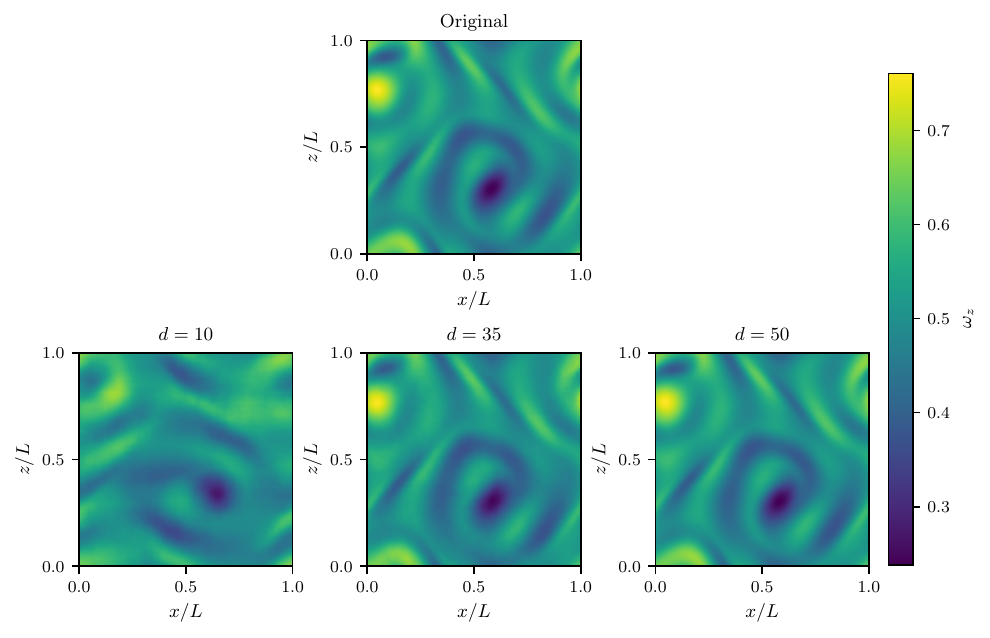}
    \caption{Original vorticity field and its reconstructions obtained from autoencoders with different latent dimensions $d$ for $\Rey=60$. The top panel shows the reference field, while the bottom panels display reconstructions for $d=10,35$ and 
$50$.}
    \label{fig:reconstructions_re60}
\end{figure}

\begin{figure}
    \centering
    \includegraphics[width=0.8\linewidth]{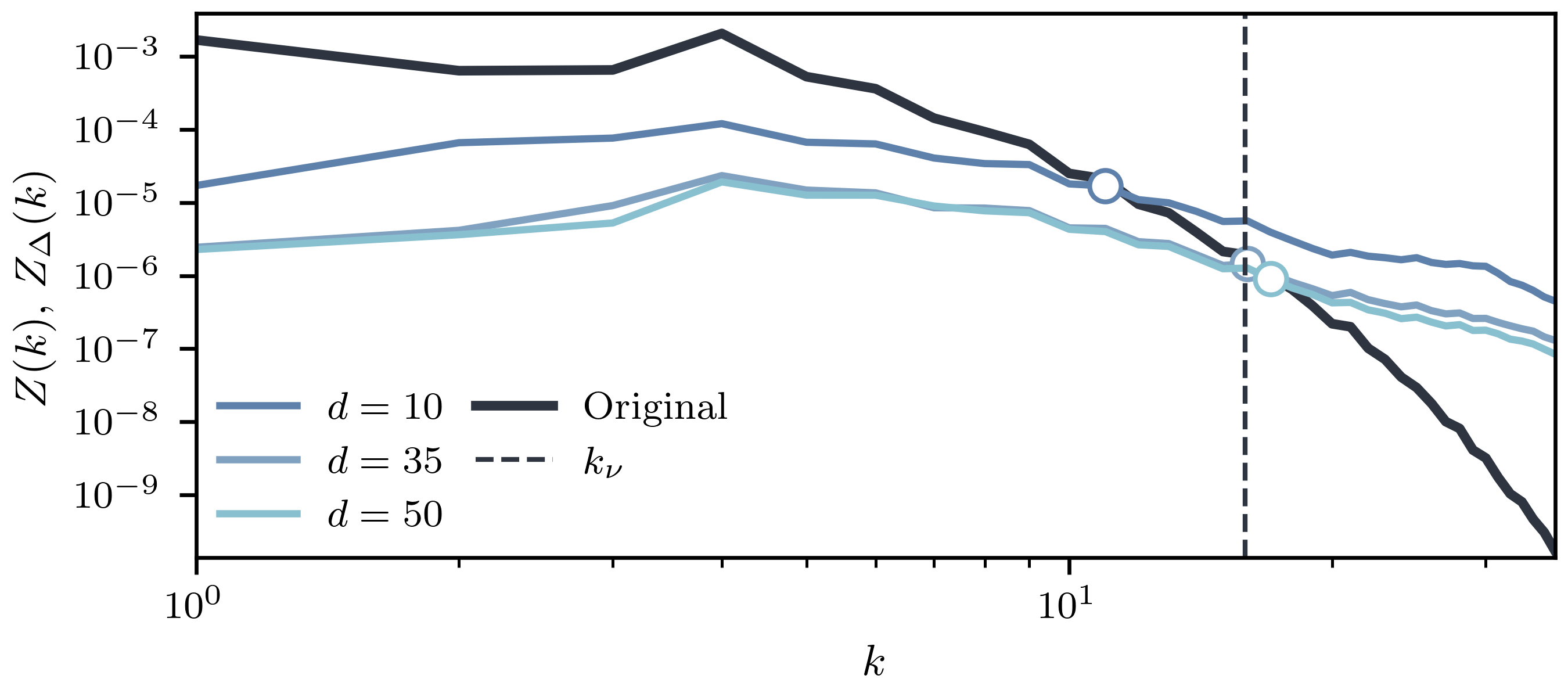}
    \caption{
    Comparison between the error vorticity spectra~$Z_\Delta(k)$ obtained for different latent dimensions~$d$ and the original vorticity spectrum~$Z(k)$ for $\Rey = 60$. 
    The spectra are computed from the test dataset, and their normalization therefore differs from that of the original simulations.
    The dots indicate the maximum resolved wavenumber~$k$ for each~$d$, and the dashed vertical line marks the small-scale dissipation wavenumber~$k_\nu$.}
    \label{fig:fig_Dk_spectra_re60}
\end{figure}

\figurename~\ref{fig:reconstructions_re60} shows representative reconstructions of the vorticity field obtained for different latent dimensions~$d$. Increasing the bottleneck size improves the reconstruction quality, yielding sharper and smaller-scale features. \figurename~\ref{fig:fig_Dk_spectra_re60} illustrates this reconstruction criterion for $\Rey=60$, presenting the corresponding spectral comparison between $Z(k)$ and $Z_\Delta(k)$, from which the maximum resolved wavenumber~$k_r(d)$ is extracted. In the spectra, a larger bottleneck layer manifests as a progressive extension of the resolved range, with $d^\ast = 35$ being the smallest dimension required to recover the flow up to the dissipation scale at $\Rey=60$, although networks with larger~$d$ (e.g.\ $d=50$) still exhibit minor improvements beyond that limit.

\bibliography{bib}

\end{document}